\documentclass[fleqn,12pt,twoside]{article}
\usepackage[headings]{espcrc1}

\readRCS
$Id: espcrc1.tex,v 1.2 2004/02/24 11:22:11 spepping Exp $
\usepackage{graphicx}
\usepackage[figuresright]{rotating}

\newcommand{\AmS}{{\protect\the\textfont2
  A\kern-.1667em\lower.5ex\hbox{M}\kern-.125emS}}

\title{Open Charm Production in $\sqrt{s_{NN}}$=200 GeV Au+Au Collisions at STAR}

\author{Haibin Zhang\address{Physics Department, Brookhaven National Laboratory
 Upton, NY 11973, USA\\Email: haibin@bnl.gov} (for the STAR\footnote[1]{For the full list of STAR authors and acknowledgements,
see appendix `Collaboration' of this volume.} Collaboration)}

\runtitle{Open Charm Production in $\sqrt{s_{NN}}$=200 GeV Au+Au
collisions at STAR} \runauthor{Haibin Zhang}

\begin{document}

\maketitle

\begin{abstract}
We present first results on $D^0$ meson production via direct
reconstruction of its hadronic decay channel $D^0\rightarrow K\pi$
in minimum bias Au+Au collisions at $\sqrt{s_{NN}}$=200 GeV with
$p_T$ up to $\sim$3 GeV/$c$. Single electron\footnote[2]{The word
``electron'' refers to electron/positron throughout these
proceedings.} spectra with 1$<p_T<$4 GeV/$c$ from the charm
semi-leptonic decays are also analyzed from the same data set. The
charm production total cross-section per nucleon-nucleon collision
is measured to be 1.11$\pm$0.08(stat.)$\pm$0.42(sys.) mb in minimum
bias Au+Au collisions, which is consistent with the $N_{bin}$
scaling compared to $d$+Au collisions. The nuclear modification
factors of the single electrons in minimum bias and 0-20\% Au+Au
collisions are significantly below unity at 1$<p_T<$4 GeV/$c$.
\end{abstract}

\section{Introduction}

Due to their large masses, charm quarks provide a unique tool to
probe the partonic matter created in relativistic heavy-ion
collisions. Charm quarks are produced in the early stages of
high-energy heavy-ion collisions~\cite{lin} and their charm total
cross-section follows the number of binary collision ($N_{bin}$)
scaling from $p+p$, $d$+Au to Au+Au collisions at RHIC energies.
Theoretical calculations~\cite{teaney,rapp} have shown that the
charm quarks interacting with surrounding partons in the medium
could change their properties, such as the $p_T$ shape, etc.
Notably, charm quarks are believed to lose less energy compared to
light quarks in the partonic matter due to the so-called
``dead-cone" effect~\cite{dead,miko,armesto}. Thus, measurements
of the $D$ mesons together with the charm-hadron decayed single
electrons - their $p_T$ distributions and nuclear modification
factors - will be vital to interpret the physics in relativistic
heavy-ion collisions.

\section{Analysis and Results}

The data used for this analysis were taken with the STAR
experiment during the $\sqrt{s_{NN}}$=200 GeV Au+Au run in 2004 at
RHIC. A minimum bias Au+Au collision trigger was defined by
requiring coincidences between two zero degree calorimeters. A
total of 13.3 and 7.6 million minimum bias triggered Au+Au
collision events are used for the $D^0$ reconstruction and the
Time-of-Flight (TOF) single electron analysis, respectively. For
the centrality dependence study of the single electrons, the
minimum bias event sample is subdivided into three centrality
bins: 0-20\%, 20-40\% and 40-80\%.

\begin{figure}[htp]
\centering
\includegraphics[height=17pc,width=30pc]{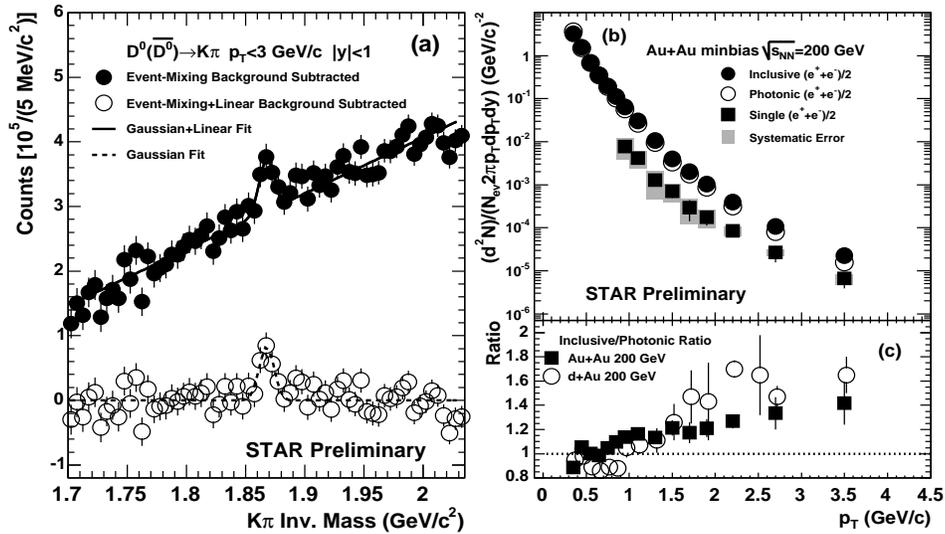}
\caption{(a) Invariant mass distributions of kaon-pion pairs after
mixed-event background subtraction (solid circles) and an
additional linear residual background subtraction (open circles).
(b) $p_T$ distributions for inclusive (solid circles), photonic
(open circles), and single (solid squares) electrons. (c) The
ratio of inclusive electrons to the photonic backgrounds in
minimum bias Au+Au (solid squares) and $d$+Au (open circles)
collisions.}\label{fig:figure1}
\end{figure}

The low $p_T$ ($<3$ GeV/$c$) $D^0$ mesons were reconstructed
through their decay $D^0\rightarrow K^-\pi^+$
($\bar{D^0}\rightarrow K^+\pi^-$) with a branching ratio of
3.83\%. Analysis details can be found in Ref.~\cite{dAuCharm}.
Panel (a) of Fig.~\ref{fig:figure1} shows the invariant mass
distributions of kaon-pion pairs after mixed-event background
subtraction (solid circles) and an additional linear residual
background subtraction (open circles). The solid circle
distributions are then fit with a Gaussian plus a linear function
to extrapolate the signal of the $D^0$ meson. After the fit, the
$D^0$ mass and width is found to be 1868$\pm$1 MeV/$c^2$ and
6$\pm$2 MeV/$c^2$, respectively. Three other $D^0$ signals are
also observed in different $p_T$ bins: 0.2$<p_T<$0.7 GeV/$c$,
0.7$<p_T<$1.2 GeV/$c$ and 1.2$<p_T<$2 GeV/$c$.

By using the combined information from the STAR TPC and TOF
detectors, electrons can be identified and measured. Detailed
analysis for the inclusive, photonic and single electron
reconstruction can be found in Ref.~\cite{dAuCharm}. Their $p_T$
spectra are shown in panel (b) of Fig.~\ref{fig:figure1}. Panel
(c) of Fig.~\ref{fig:figure1} shows the ratio of inclusive
electrons to the photonic backgrounds as a function of $p_T$. A
significant excess is observed at $p_T>$1 GeV/$c$ that is expected
to be dominated by electrons from semi-leptonic decays of
charm-hadrons.

\begin{figure}[htp]
\centering
\includegraphics[height=17pc,width=30pc]{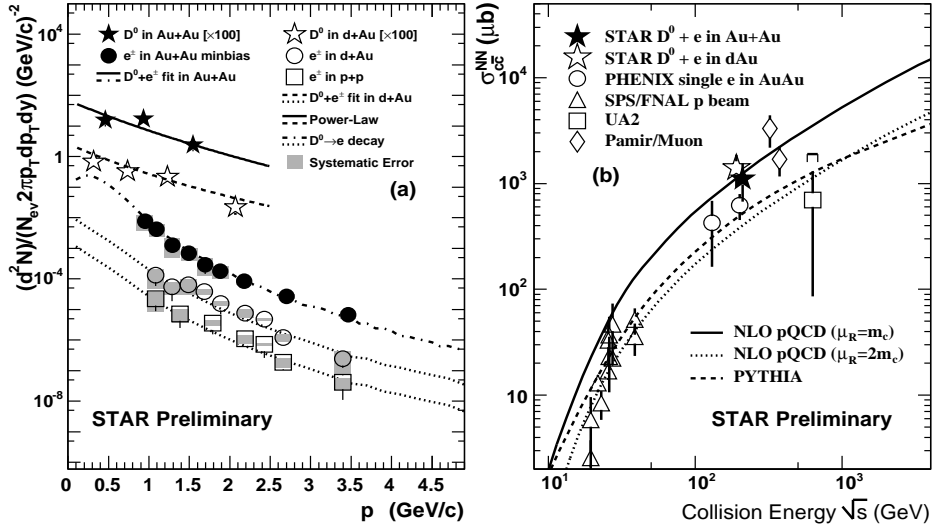}
\caption{(a) $p_T$ distributions of $D^0$ in Au+Au (filled stars)
and $d$+Au (open stars) collisions and single electrons in Au+Au
(solid circles), $d$+Au (open circles) and $p+p$ (open squares)
collisions. (b) The $c\bar{c}$ cross-section per nucleon-nucleon
collision vs. the collision energy.}\label{fig:figure2}
\end{figure}

Panel (a) of Fig.~\ref{fig:figure2} shows the $p_T$ distributions of
$D^0$ and single electrons in minimum bias Au+Au, $d$+Au and $p+p$
collisions at $\sqrt{s_{NN}}$=200 GeV. Using a combined fit applied
to both, directly reconstructed $D^0$ and the single electron
distribution in Au+Au collisions, the mid-rapidity $D^0$ yield is
then obtained and converted to the charm total cross-section per
nucleon-nucleon collision ($\sigma_{c\bar{c}}^{NN}$) following the
method addressed in Ref.~\cite{dAuCharm}. $\sigma_{c\bar{c}}^{NN}$
is measured to be 1.11$\pm$0.08(stat.)$\pm$0.42(sys.) mb from
minimum bias Au+Au collisions which is comparable to
$\sigma_{c\bar{c}}^{NN}$=1.4$\pm$0.2$\pm$0.2 mb in minimum bias
$d$+Au collisions at $\sqrt{s_{NN}}$=200 GeV. Thus the charm total
cross-section roughly follows the $N_{bin}$ scaling from $d$+Au to
Au+Au collisions which supports the conjecture that charm quarks are
produced at early stages in relativistic heavy-ion collisions.

\begin{figure}[htp]
\centering
\includegraphics[height=17pc,width=30pc]{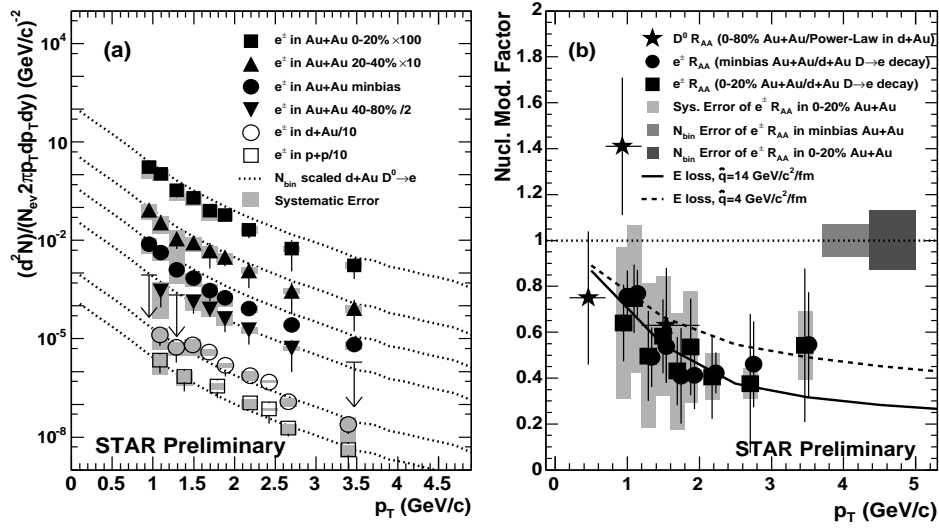}
\caption{(a) Single electron $p_T$ distributions from minimum bias
Au+Au (solid circles), 0-20\% (solid squares), 20-40\% (upward
triangles) and 40-80\% (downward triangles) centralities of Au+Au
collisions, $d$+Au (open circles) and $p+p$ (open squares)
collisions. (b) Nuclear modification factors for $D^0$ (stars) and
single electrons (circles and squares). Solid and dashed curves
are theoretical predictions for single electron $R_{AA}$
in~\cite{armesto} with different parameter
settings.}\label{fig:figure3}
\end{figure}

The $D^0$ $R_{AA}$ (star symbols in Panel (b) of
Fig.~\ref{fig:figure3}) are calculated by dividing the $D^0$ data
points in minimum bias Au+Au collisions by the power-law fit
results of the $D^0$ $p_T$ spectrum in $d$+Au collisions scaled by
$N_{bin}$. The single electron $p_T$ spectra are also measured in
three collision centralities in Au+Au collisions: 0-20\%, 20-40\%
and 40-80\%, shown in Panel (a) of Fig.~\ref{fig:figure3} and
compared to the $D^0\rightarrow e^{\pm}$ decayed shape in $d$+Au
collisions scaled by $N_{bin}$. The spectra in minimum bias and
0-20\% Au+Au collisions significantly deviate from the curves. The
single electron $R_{AA}$ in minimum bias and 0-20\% central Au+Au
collisions can then be calculated by dividing the corresponding
data points by their referring curves, shown as solid circles and
squares in Panel (b) of Fig.~\ref{fig:figure3}, respectively. The
single electron $R_{AA}$ in both minimum bias and 0-20\% central
Au+Au collisions is observed to be significantly below unity at
1$<p_T<$4 GeV/$c$ and is consistent with theoretical predictions
in~\cite{armesto} by considering charm quark radiation energy loss
with $\hat{q}$=14 GeV/$c^2$/fm indicating that the charm $p_T$
spectra must have been modified by the hot and dense medium in
relativistic heavy-ion collisions.

\section{Conclusions}

In conclusion, the $D^0$ and the non-photonic single electron
$p_T$ distributions are measured in Au+Au collisions at
$\sqrt{s_{NN}}$=200 GeV at STAR. Within the present statistical
and systematic errors, the charm total cross-section per
nucleon-nucleon collision is consistent with $N_{bin}$ scaling
from $d$+Au to Au+Au collisions indicating   the charm quarks are
mostly produced in early stages at relativistic heavy-ion
collisions. The single electron nuclear modification factors in
0-20\% centrality and minimum bias Au+Au collisions are measured
to be significantly below unity at 1$<p_T<$4 GeV/$c$. Therefore we
conclude that the charm $p_T$ spectra are indeed modified by the
hot and dense medium in relativistic heavy-ion collisions at RHIC
energies.


\begin{thebibliography}{9}
\bibitem{lin} Z. Lin et al. Phys. Rev. C 51 (1995) 2177.
\bibitem{teaney} G. Moore et al. Phys. Rev. C 71 (2005) 064904.
\bibitem{rapp} H. van Hees et al. nucl-th/0508055.
\bibitem{dead} Y. Dokshizer et al. Phys. Lett. B 519 (2001) 199.
\bibitem{miko} M. Djordjevic et al. Phys. Rev. Lett. 94 (2005) 112301.
\bibitem{armesto} N. Armesto et al. Phys. Rev. D 71 (2005) 054027.
\bibitem{dAuCharm} J. Adams et al. Phys. Rev. Lett. 94 (2005) 062301.
\end{thebibliography}
\end{document}